\documentstyle[11pt]{article}
\begin{document}
\begin{center}
{\Large\bf Chiral Symmetry and Electroweak $\pi ,K,\eta$ Processes}\\

\medskip

Barry R. Holstein\\
Department of Physics and Astronomy\\
University of Massachusetts\\
Amherst, MA 01003 \\

\medskip
\end{center}
\medskip
\begin{abstract}
Recently, the development of chiral perturbation theory has allowed the
generation of rigorous low-energy theorems for various hadronic processes
based only on the chiral invariance of the underlying QCD Lagrangian.
Herein we examine the experimental implications of chiral symmetry in
the regime of electroweak Goldstone boson interactions.

\end{abstract}

\section{Introduction}
It has long been the holy grail for particle and nuclear knights
to generate rigorous
predictions from the Lagrangian of QCD
\begin{equation}
{\cal L}_{\rm QCD}=-{1\over 2}G_{\mu\nu}G^{\mu\nu}+\bar{q}(i\gamma_\mu
D^{\mu}-m)q.
\end{equation}
Despite the ease with which one can write this equation,
because of its inherent nonlinearity progress in this regard has been
slow.   In recent years, however,  procedure has been developed---chiral
perturbation theory ($\chi$PT)---which exploits the (broken) chiral symmetry
of QCD and
allows rigorous predictive power in the case of low energy reactions.  This
technique, based on a suggestion due to Weinberg\cite{1} was developed
(at one loop level) during the last decade in an important series of papers by
Gasser
and Leutwyler and others,\cite{2} and is based on the feature that the QCD
Lagrangian (Eq. 1) has a global $SU(3)_L\times SU(3)_R$ (chiral)
invariance in the limit of vanishing quark mass.  Such invariance is manifested
in the real world not in the conventional fashion but rather is spontaneously
broken, resulting in eight light Goldstone bosons---$\pi, K,\eta$---which
would be massless if the corresponding quark masses also vanished.
While the identification of this symmetry is apparent in terms of quark/gluon
degrees of
freedom, it is not so simple to understand the implications of chiral
invariance
in the arena of experimental meson/baryon interactions.

Early attempts in this direction were based on current algebra/PCAC
methods, yielding relationships between processes differing in the
number of pions, e.g.
\begin{equation}
\lim_{q\rightarrow 0}<B\pi^a_q|{\cal O}|A>={-i\over F_\pi}<B|[F_5^a,{\cal
O}]|A>
\end{equation}
where $F_\pi =92.4 MeV$ is the pion decay constant.  However, recently
we have learned how to study chiral strictures
using so-called effective Lagrangian techniques,
and the development of $\chi$PT has opened up an important window on
the low energy interactions of these Goldstone particles which has heretofore
been unavailable and which succinctly expresses all information about such
reactions in the energy regime $E<\sim 0.6GeV$ in terms of just ten
phenomenological constants $L_1,\ldots L_{10}$.  Because of space limitations
we shall not be able here to outline this chiral technology but rather refer
the
interested reader to the relevant literature.  We shall have to be satisfied
 in the next sections to outline the results of such calculations within the
electroweak interactions of $\pi ,K,\eta$ mesons respectively, indicating
where possible problems and challenges lie for future experimental and/or
theoretical work.

\section{$\pi ,K$ Reactions}

The simplest pionic process for which chiral perturbation theory makes a
prediction is that of radiative pion
decay---$\pi^+\rightarrow e^+\nu_e\gamma$---for which the decay amplitude
assumes the form
\begin{eqnarray}
M_{\mu\nu}(p,q)&=&\int d^4x e^{iq\cdot x}<0|T(V_\mu^{\rm em}(x)J_\nu^{\rm wk}
(0))|\pi^+({\bf p})>={\rm Pole}\nonumber\\
& &+h_A[(p-q)_\mu q_\nu -g_{\mu\nu}(p-q)\cdot q]+r_A(q_\mu q_\nu
-g_{\mu\nu}q^2)
+ih_V\epsilon_{\mu\nu\alpha\beta}p^\alpha q^\beta
\end{eqnarray}
In general then there exist two axial ($h_A,r_A$) and one vector ($h_V$)
structure functions.  However, there is a catch.
Chiral symmetry does {\it not} predict the size of $h_A$, for which a rather
precise number is available.   Rather this parameter is used as input
for determination of one of the GL parameters---$L_{10}$.
In order to
determine $h_V$ and $r_A$ the rather rare
($\sim 10^{-9}$) Dalitz mode---$\pi^+\rightarrow e^+\nu_ee^+e^-$---must be
employed and the limits obtained thereby are somewhat imprecise\cite{3}
\begin{eqnarray}
h_V|^{\rm theo}&=&{1\over 4\sqrt{2}\pi^2F_\pi}=0.027m_\pi^{-1}\quad
{\rm vs.}\quad h_V|^{\rm exp}=(0.029\pm 0.017)m_\pi^{-1}\nonumber\\
{r_A\over h_V}|^{\rm theo}&=&{8\pi^2F_\pi^2\over 3}<r_\pi^2>\simeq
2.6\quad{\rm vs.}\quad{r_A\over h_V}|^{\rm exp}=2.3\pm 0.6
\end{eqnarray}
Thus while both numbers are in agreement with the chiral restrictions there
is also plenty of room for improvement in experimental precision.

An important probe of pion structure is provided by of its
electric (magnetic) polarizability $\alpha_E (\beta_M)$,
which measures the constant of
proportionality between induced dipole moments and applied electric
(magnetic) fields.\cite{4}  The polarizability may be probed via the Compton
scattering process, which to lowest order must assume the form
\begin{equation}
T_{\rm Compton}=\hat{\epsilon}\cdot\hat{\epsilon}'({-Q^2\over
m}+\omega\omega'4\pi\alpha_E)
+\hat{\epsilon}\times{\bf q}\cdot\hat{\epsilon}'\times{\bf
q}'4\pi\beta_M+\ldots
\end{equation}
Chiral symmetry predicts the size of both electric and magnetic
polarizabilities
in terms of the axial structure function $h_A$ via\cite{5}
\begin{equation}
\alpha_E=-\beta_M={\alpha\over 8\pi^2m_\pi F_\pi^2}{h_A\over h_V}=
(2.8\pm 0.3)\times 10^{-4}{\rm fm}^3
\end{equation}
The sum of electric and magnetic polarizabilities is predicted to vanish,
in agreement with experiment---
\begin{equation}
\alpha_E+\beta_M=(1.4\pm 3.1)\times 10^{-4}{\rm fm}^3\cite{6}
\end{equation}
but there is not yet agreement on the experimental size of the electric
polarizability, which has been measured in three different fashions
\begin{eqnarray}
\alpha_E  &=& (6.8\pm1.4)\times 10^{-4}{\rm fm}^3\quad (\hbox
{radiative pion scattering})\cite{7}
\nonumber\\
 &=& (20\pm 12)\times 10^{-4}{\rm fm}^3\quad (\hbox {radiative pion
production})\cite{8}\nonumber\\
 &=& (2.2\pm 1.6)\times 10^{-4}{\rm fm}^3 \quad (\gamma\gamma\rightarrow\pi\pi
)
\cite{9}
\end{eqnarray}
Clearly there exists
a lack of agreement here and clarifying experimental work is
called for on this very fundamental aspect of the pion.

Moving the the kaon sector, there is a complete correspondence between
pion quantities and their kaonic analogs, and the latter are completely
predicted by chiral symmetry---
\begin{equation}
h_A^K=h_A^\pi ,\quad r_A^K=r_A^\pi ,\quad h_V^K=h_V^\pi ,\quad\alpha_E^K=
-\beta_M^K={m_\pi F_\pi^2\over m_KF_K^2}\alpha_E^\pi.
\end{equation}
However, the experimental information is very limited and is restricted
to the sum of vector and axial structure functions in radiative kaon decay
\begin{equation}
h_A^K+h_V^K|^{\rm exp}=(0.043\pm 0.003)m_\pi^{-1}\cite{10}\quad
{\rm vs.}\quad h_A^K+h_V^K|^{\rm theo}=0.038m_\pi^{-1}.
\end{equation}
The existence of a high intensity kaon factory could clearly have a significant
impact in this regard.

Although there is much more which we could discuss such as
$K_{\ell 3},K_{\ell 4}$ decays and their radiative partners, for space reasons
we move on to consider the nonleptonic kaon sector, where chiral symmetry
also is a powerful tool.  Besides the dominant modes $K\rightarrow 2\pi ,3\pi$
which are closely related via soft pion theorems, there is special interest
in the nonleptonic-radiative processes $K\rightarrow \gamma\gamma ,
\pi^0\gamma\gamma$, which occur at one loop order in the chiral expansion.
In the former case one finds a prediction\cite{14}
\begin{eqnarray}
\Gamma (K_S\rightarrow\gamma\gamma)&=&{\alpha^2m_K^3g_8^2F_\pi^2\over 16\pi^3}
(1-{m_\pi^2\over m_K^2})|F({m_K^2\over m_\pi^2})|^2\nonumber\\
{\rm where}\quad F(z)&=&1-{1\over z}\ln^2 ({\beta (z) +1\over \beta (z) -1})
\end{eqnarray}
with $\beta =\sqrt{1-{4\over z}}$ being the pion velocity in $K_S\rightarrow
\pi\pi$ and $g_8 \approx 7.8\times 10^{-8}F_\pi^2$ is a parameter
determined via the
$K_S\rightarrow \pi\pi$
decay rate.  The branching ratio predicted in this way is in good agreement
with the value recently measured at CERN\cite{12}
\begin{equation}
B(K_S\rightarrow \gamma\gamma )|^{\rm theo}=2.0\times 10^{-6} \quad
{\rm vs.}\quad B(K_S\rightarrow\gamma\gamma )|^{\rm exp}=(2.4\pm 1.2)\times
10^{-6}.
\end{equation}
Not so simple, on tthe other hand, is the process $K_L\rightarrow
\pi^0\gamma\gamma$
for which we predict
\begin{eqnarray}
{d\Gamma\over dz}(K_L\rightarrow \pi^0\gamma\gamma )&=&{\alpha^2m_K^5g_8^2\over
(4\pi )^5}\lambda^{1\over 2}(1,z,{m_\pi^2\over m_K^2})
 |(z-{m_\pi^2\over m_K^2})F(z{m_K^2\over m_\pi^2})+(1-z+{m_\pi^2\over
m_K^2})F(z)|^2\nonumber\\
{\rm with}\quad z&=&{m_{\gamma\gamma}^2\over m_K^2}, \quad\lambda(a,b,c)
=a^2+b^2+c^2-2(ab+ac+bc)
\end{eqnarray}
The expected spectrum is very distinctive, with most of the events
predicted to occur in the high z range and this is in agreement with present
experimental indications.  However, the predicted branching ratio
is {\it not}
\begin{equation}
B(K_L\rightarrow\pi^0\gamma\gamma )|^{\rm theo}=6.8\times 10^{-7}
\quad {\rm vs.}\quad B(K_L\rightarrow\pi^0\gamma\gamma )|^{\rm exp}=
(2.0\pm 0.5)\times 10^{-6}\cite{13}
\end{equation}
indicating the need for additional experimental and theoretical effort.
While we could continue this discussion of the nonleptonic kaon sector
considerably, space limitations require that we move now to
the realm of eta decay.

\section{Eta Decay Processes}

The challenge of dealing with decay of $\eta (547)$ involves
inclusion of the mixing with
$\eta '(958)$, which lies outside the simple chiral $SU(3)_L\times SU(3)_R$
framework.  To lowest order things are simple---in the chiral limit the
pseudoscalar mass spectrum would consist of a massless octet of Goldstone
bosons plus a massive $SU(3)$ singlet ($\eta_0$).  With the breaking of
chiral invariance the octet pseudoscalar masses become nonzero, and are
related at first order in symmetry breaking by the Gell-Mann-Okubo formula
\begin{equation}
m^2_{\eta_8}={4\over 3}m_K^2-{1\over 3}m_\pi^2
\end{equation}
where $\eta_8$ is the eighth member of the octet.  At this same order in
symmetry breaking the singlet $\eta_0$ will in general mix with $\eta_8$
producing the physical eigenstates $\eta,\eta '$ given by
\begin{equation}
\eta =\cos\theta \eta_8-\sin\theta\eta_0,\quad
\eta '=\sin\theta\eta_8+\cos\theta\eta_0.
\end{equation}
The mixing angle $\theta$ can be determined via diagonalization of the mass
matrix
\begin{equation}
m^2=\left( \begin{array}{cc}
   m_{\eta_8}^2 & \kappa \\
   \kappa       & m_{\eta_0}^2 \\
\end{array} \right)
\end{equation}
Taking $m_{\eta_8}$ from Eq. 15 and fitting $m_{\eta_0},\kappa$ with the two
known masses yields the prediction $\theta = -9.4^\circ $.
However, there is good reason not to trust this traditional analysis, since
higher order chiral symmetry breaking terms can make important modifications.
For example, inclusion of the leading chiral log correction from meson-meson
scattering, we find\cite{14}
\begin{equation}
m_{\eta_8}^2={4\over 3}m_K^2-{1\over 3}m_\pi^2-{2\over 3}{m_K^2\over (4\pi
F_\pi )^2}{\rm ln}{m_K^2\over \mu^2}
\end{equation}
for which diagonalization of the mass matrix yields
$\theta \approx -19.5^\circ .$
Of course, this is just an approximate result.  However, a full one loop
calculation using $\chi$PT yields essentially the same value.\cite{2}
At this same (one-loop) level of symmetry breaking there is generated a shift
in the lowest order value of the pseudoscalar decay constant $F_P$---
\begin{eqnarray}
F_\pi &=&\bar{F}\left[1-{1\over 2}{m_K^2\over (4\pi F_\pi)^2}\ln {m_K^2\over
\mu^2}\right] \approx 1.12\bar{F}\nonumber\\
F_{\eta_8}&=&\bar{F}\left[1-{3\over 2}{m_K^2\over (4\pi F_\pi )^2}\ln
{m_K^2\over
\mu^2}\right] \approx 1.25 F_\pi \quad {\rm for} \quad \mu\approx 1 GeV.
\end{eqnarray}

With this introductory material in hand we can now confront the
remaining subject of
our report---that of eta decay.  First consider the dominant two-photon
decay mode, which to leading order arises due to the anomaly.  In the
analogous $\pi^0\rightarrow \gamma\gamma$ case we find
\begin{equation}
{\rm Amp}\equiv F_{\pi\gamma\gamma}(0)\epsilon^{\mu\nu\alpha\beta}\epsilon_\mu
k_\nu\epsilon'_\alpha k'_\beta \quad{\rm with}\quad F_{\pi\gamma\gamma}(0)=
{N_c\alpha\over 3\pi F_\pi }=0.025 GeV^{-1}.
\end{equation}
General theorems guarantee that this result is not altered in higher orders
of chiral symmetry breaking and the experimental value\cite{14}
\begin{equation}
F_{\pi\gamma\gamma}=(0.0250\pm 0.0005) GeV^{-1}
\end{equation}
is in excellent agreement with its theoretical analog, eloquently
confirming the value $N_c=3$ as the number of colors.
The $\eta ,\eta '\rightarrow \gamma\gamma$ couplings
also arise from the anomalous
component of the effective chiral Lagrangian, and in an extended $\chi$PT
approximation have the values
\begin{eqnarray}
F_{\eta\gamma\gamma}(0)&=&{F_{\pi\gamma\gamma}(0)\over \sqrt{3}}
\left({F_\pi\over F_8}\cos\theta
-2\sqrt{2}{F_\pi\over F_0}\sin\theta\right)\nonumber\\
F_{\eta '\gamma\gamma}(0)&=&{2\sqrt{2}F_{\pi\gamma\gamma}(0)\over \sqrt{3}}
\left( {1\over 2\sqrt{2}}{F_\pi\over F_8}\sin\theta + {F_\pi\over F_0}\cos
\theta )\right) .
\end{eqnarray}
The experimental numbers
\begin{equation}
F_{\eta\gamma\gamma}(0)=0.0249\pm 0.0010 GeV^{-1} \quad
F_{\eta'\gamma\gamma}(0)=0.0328\pm 0.0024 GeV^{-1}
\end{equation}
are fit well by
$F_8/ F_\pi \approx 1.24$  and $F_0/  F_\pi\approx 1.04.$
Note also that the value of $F_8 /F_\pi $ is in good agreement with that
expected
from chiral arguments given above.

Processes involving a photon coupled to three pseudoscalar mesons also
involve the
anomaly and at zero four-momentum are completely determined.
First consider
the case of $\gamma\rightarrow\pi^+\pi^-\pi^0$.  At zero four-momentum the
anomaly requires\cite{15}
\begin{eqnarray}
{\rm Amp}(3\pi -\gamma)&=&A(s_{+-},s_{+0},s_{-0})\epsilon^{\mu\nu\alpha\beta}
\epsilon_\mu p_{+\nu}p_{-\alpha}p_{0\beta}\nonumber\\
{\rm where}\quad A(0,0,0)&=&{eN_c\over 12 \pi^2F_\pi^3}=9.7GeV^{-3}
\quad{\rm and}\quad s_{ij}=(p_i+p_j)^2
\end{eqnarray}
Inclusion of additional diagrams yields
\begin{equation}
A(s,t,u)={eN_c\over 12\pi^2F_\pi^3}\left[1+{1\over 2}\left({s\over m_\rho^2-s}
+{t\over m_\rho^2-t}+{u\over m_\rho^2-u}\right)\right]
\end{equation}
which has the structure required by vector dominance, and agrees with the
value required by the chiral anomaly at zero four-momentum.
 The $\gamma -3\pi$ reaction has been studied experimentally via
pion pair production by the pion in the nuclear Coulomb field
and yields a number\cite{16}
\begin{equation}
A(0,0,0)_{\rm exp}=12.9\pm 0.9 \pm 0.5 GeV^{-3}
\end{equation}
in apparent disagreement with Eq. 24 and suggesting the
value $N_c\approx 4$!  The most likely conclusion is that this an experimental
problem associated with this difficult-to-measure process, but in any
case a new high-precision
experiment would be of great interest.

Having warmed up on the $\gamma -3\pi$ process, it is now
straightforward to construct the analogous $\eta\rightarrow\pi^+\pi^-\gamma$
amplitude, for which we find
\begin{equation}
{\rm Amp}(\eta\rightarrow\pi^+\pi^-\gamma )=B(s_{+-},s_{+\gamma},s_{-\gamma})
\epsilon^{\mu\nu\alpha\beta}\epsilon^*_\mu p_{+\nu}p_{-\alpha}k_{\gamma\beta}
\end{equation}
with
\begin{eqnarray}
& &B(s,t,u)= B(0,0,0)
\times \left[ 1+{3\over 2}{s\over m_\rho^2-s }\right] \qquad {\rm and}
\nonumber\\
& &B(0,0,0)={eN_c\over 12\sqrt{3}\pi^2F_\pi^3 }\left( {F_\pi\over
F_8}\cos\theta
-\sqrt{2}{F_\pi \over F_0}\sin\theta \right)= 6.81 GeV^{-3}
\end{eqnarray}
The $\eta\rightarrow\pi^+\pi^-\gamma$ reaction was studied in the experiment
of Layter et al.\cite{17} and yielded
\begin{equation}
|B(0,0,0)|_{\rm exp}=(6.47\pm 0.25) GeV^{-3}
\end{equation}
which is in reasonable agreement with Eq. 28 and
reinforces the validity of the numbers obtained in the two photon
analysis.

Having above confirmed the basic correctness of the predictions of the anomaly
(and thereby of this important cornerstone of QCD) we move now to the important
three pion decay of the eta, which
rather probes the {\it conventional} two- and four-derivative piece of the
chiral
Lagrangian.  The decay of the isoscalar eta to the predominantly I=1 final
state of the
three pion system occurs primarily due to the u-d quark mass difference,
and the result arising from lowest order
chiral perturbation theory is well-known
\begin{equation}
{\rm Amp}(\eta_8\rightarrow\pi^+\pi^-\pi^0)={-B_0(m_d-m_u)\over
3\sqrt{3}F_\pi^2}\left[1+{3(s-s_0)\over m_\eta^2-m_\pi^2}\right].
\end{equation}
The d-u quark mass difference has traditionally been extracted from the
experimental $K^+-K^0$ mass splitting with the electromagnetic component
eliminated via use of Dashen's theorem---\cite{18}
\begin{equation}
(m_{\pi^+}^2-m_{\pi^0}^2)=(m_{K^+}^2-m_{K^0}^2)_{\rm EM}.
\end{equation}
This assumption results in a prediction in serious contradiction to the
experimental result
\begin{equation}
\Gamma (\eta\rightarrow\pi^+\pi^-\pi^0)^{\rm theo}=66eV\quad {\rm vs.}\quad
\Gamma (\eta\rightarrow\pi^+\pi^-\pi^0)_{\rm exp}=310\pm  50eV.
\end{equation}
At first sight this would appear to be a rather strong and irreparable
violation of a lowest order chiral prediction and therefore not salvagable
by the expected ${\cal O}(m_\eta^2/(4\pi F_\pi^2)^2\sim 30\% $ corrections
from higher order effects.  However, this is not at all the case.
The one-loop and counterterm contributions were calculated by Gasser
and Leutwyler and were found to enhance the lowest order prediction by a
factor 2.6, and recent work has suggested a significant violation of
Dashen's theorem---
\begin{equation}
(m_d-m_u)_{\chi -{\rm broken}}\approx 1.2(m_d-m_u)_{\rm Dashen}\cite{19}
\end{equation}
which corresponds to an additional 40\% enhancement of the chiral estimate,
{\it i.e.} $\Gamma (\eta\rightarrow\pi^+\pi^-\pi^0)\sim 240 eV$,
and puts the result now in the right ballpark.

In order to decide the origin of any remaining discrepancy, it is necessary
to make careful spectral shape measurements.  Phenomenologically, we expand the
decay amplitude about the center of the Dalitz plot as
\begin{equation}
{\rm Amp}\equiv \alpha\left[1 +\beta Y +\gamma (Y^2+{1\over 3}
X^2)+\delta (Y^2-{1\over 3}X^2)\right]
\end{equation}
where $X,Y$ are the usual Dalitz variables.
These parameters have been determined phenomenologically
to be\cite{20}
\begin{eqnarray}
\beta &=& 0.216\pm 0.003\qquad\gamma = -0.0067\pm 0.003\qquad\delta
=-0.0139\pm 0.003\nonumber\\
\beta&=& 0.234\pm 0.004\qquad \gamma = -0.0006\pm 0.003\qquad\delta
=-0.0099\pm 0.003\nonumber\\
& &\hfill
\end{eqnarray}
to be compared to the one-loop chiral prediction
\begin{equation}
\beta = 0.266 \qquad\gamma =  0.0054 \qquad\delta = -0.0072
\end{equation}
We see that there is general though certainly not excellent agreement.

An additional test of the validity of the chiral approach lies in our ability
to predict the $\eta\rightarrow 3\pi^0$ reaction, for which one finds
\begin{equation}
{\Gamma (\eta\rightarrow \pi^+\pi^-\pi^0)\over
\Gamma (\eta\rightarrow \pi^0\pi^0\pi^0)}
=\left\{ \begin{array}{ll}
1.5 & {\cal O}(p^2) \\
1.43 & {\cal O}(p^4) \\
1.3 & {\rm experiment} \\
\end{array} \right.
\end{equation}
Clearly there is plenty of challenge---both theoretical and experimental---
in the eta system.

\section{Conclusions}

We have seen above that chiral symmetry provides an important link between
experimental low enrgy physics within the Goldstone boson sector and the
QCD Lagrangian which presumably underlies it.  Despite the evident
success of such methods, however, a number of challenges remain.  These
include i) pions: clearing up apparent discrepancies in the anomalous
process $\gamma\rightarrow 3\pi$ and in the charged pion polarizability;
ii) kaons: providing experimental numbers which are presently unmeasured
in radiative semileptonic decay and clarifying the
nonleptonic-radiative $K_L\rightarrow\pi^0\gamma\gamma$ process;
iii) etas: a precise
measurement of the eta lifetime would be of interest in that present
values obtained via different methods disagree, and a new and more
precise experiment on the $\eta\rightarrow 3\pi$ spectrum would be very
useful as a test of chiral methods.  In summary, there is plenty
of interesting physics here for experimentalists and theorists alike.

\medskip

{\bf Acknowledgements}:  We thank John Donoghue for many clarifying
discussions.
This work was supported in part by the National Science Foundation.

\end{document}